\documentclass[%
 reprint,
 amsmath,amssymb,
aps,
]{revtex4-1}

\usepackage{graphicx}
\usepackage{dcolumn}
\usepackage{bm}
\usepackage{epstopdf}
\usepackage{mathtools}
\usepackage{amsmath}
\usepackage{array}
\usepackage{amsmath,amssymb}
\usepackage{lipsum}
\usepackage{subcaption}
\usepackage{CJK}

\begin{document}


\title{Phase Synchronization of Stimulated Raman Process in Optical Fiber For Long Pulse Regime}

\author{Partha Mondal}
 \altaffiliation[ ]{Department of Physics, Indian Institute of Technology, Kharagpur - 721302, India, Email: parthaphotonica@gmail.com}
 
\author{Shailendra K. Varshney}%

\altaffiliation[ ]{Department of E $\&$ ECE, Indian Institute of Technology, Kharagpur - 721302, India}

\date{\today}

\begin{abstract}
We investigate the evolution of coherence property of noise-seeded Stokes wave in short ($<$ 1 ps) and long pulse ($>$ 1 ps) regimes. Nonlinear equations expressing the evolution of pump and Stokes wave are solved numerically for both the regions. The simulations include quantum noise by incorporating noise seed in the pump field where one photon per mode with random phase. The spectral phase fluctuations of the Stokes wave for both the regions, are characterized by performing multiple simulations and finally,  the degrees of first-order mutual coherence are calculated as a function of wavelength for different conditions. Our statistical analysis proclaim that noise-seeded stimulated Raman process, which plays the role in degradation of coherence in short pulse region, exhibits strong phase synchronization in long pulse regime. The manifestation of phase synchronization occurs by the transition of the Stokes wave from incoherent to coherent spectra in long pulse regime. 
\end{abstract}

\maketitle
\section{Introduction}
Synchronization behavior is an omnipresent natural phenomena comes out in physical, chemical and biological systems \cite{synchronization_coupled_oscillation, strogatz_book}. Large amount of theoretical and experimental studies have been accomplished to understand various aspect of synchronization in classical system like flashing fireflies \cite{flashing_fireflies}, Huygen’s clocks \cite{Hygens_clock}, pacemaker cells in mammalian hearts \cite{pacemakers_cell}, oscillations of interacting Josephson junctions \cite{josephson_junction1,josephson_junction2,josephson_junction3} etc., as well as in quantum system like cold atom systems \cite{cold_atom1,cold_atom2,cold_atom3}, nanomechanical resonators \cite{nanomechanical_synch_1,nanomechanical_synch_2,nanomechanical_synch_3}, spintronics and so on. More recently, the idea of synchronization has been extended to optical system where different mathematical models like Kuramoto model \cite{frequency_comb1}, reduced phase model \cite{phase_reduction_model1,phase_reduction_model2}, have been employed to understand different nonlinear phenomena such as phase locking \cite{phase_locking1,phase_locking2}, optical frequency comb \cite{frequency_comb2,frequency_comb3,frequency_comb4,frequency_comb5} and coherent beam combining \cite{coherent_beam_combing1}.

Researchers have harnessed nonlinear interactions through Kerr medium and utilize it to realize myriad of applications in several fields. Stimulated Raman scattering (SRS) is an important nonlinear phenomena where an intense pump pulse propagating through Raman-active medium, generates downshifted Stokes wave. It was first observed by E. J. Woodbury \textit{et. al.} in 1962 \cite{1st_raman}. The Stokes wave builds up from spontaneously scattered Raman-shifted photon accompanied by amplification during the propagation of pump pulse through the Raman medium. A full description of quantum mechanical approach to study the build up of Stokes wave through SRS was developed which unifies spontaneous Raman scattering and the effect of spatial propagation through the medium \cite{SRS_unified_theory}. The theoretical study based on this model has  largely divided the growth of Stokes wave into two different regions based on the temporal duration of the pump pulses: (a) Transient region where pump pulse duration is short comparable to the molecular relaxation time i,e, $\Gamma\tau \leq (gz)^{-1}$, where $\Gamma$ is the dephasing rate and $\tau$ is the pump pulse duration and $g$ is the Raman gain coefficient (b) Steady-state region which is achieved for longer pulse width when $\Gamma\tau\geq gz$. Many theoretical and experimental works have been carried out to explore the characteristics of Stokes wave for both the regions \cite{SRS_transient_regime1,SRS_transient_regime2,SRS_transient_regime3}. The intrinsic quantum noise associated with spontaneous scattering introduces pulse to pulse fluctuations in amplitude and phase of the generated Stokes wave. Detailed study of spatial and temporal coherence property of the generated Stokes wave was adopted considering the three dimensional propagation and collisional dephasing of the propagating medium \cite{coherence_stokes1,coherence_stokes2,coherence_stokes3}. The relative phase of two Stokes pulses generated through the set of hydrogen molecules was measured interferometrically and was detected correlation for time delays upto nine collisional dephasing times \cite{memory_stokes}. Several attempt have been performed to achieve coherent Stokes wave in gas-filled hollow-core photonic crystal fiber (HC-PCF) by controlling the transient regimes and steady-state regimes of SRS amplification \cite{HC_PCF1,HC_PCF2,HC_PCF3,HC_PCF5}. First experimental observation of SRS in glass was reported by Alfano
and Shapiro in 1970 \cite{1st_SRS_glass}. A theoretical model was developed to understand the formation of Stokes wave for long pulse regime in silica core optical fiber neglecting the effect of self-phase modulation (SPM), cross-phase modulation (XPM) and group-velocity dispersion (GVD) \cite{SRS_silica_core_fiber}. Later on, a realistic model to study the growth of Stokes wave from spontaneous noise through SRS in optical fiber for long pulse regime was developed which incorporates the effect of pump depletion, GVD, SPM and XPM \cite{Raman_generation2}. Numerical simulations also have been employed in this work to investigate the shot-to-shot pulse width and energy fluctuation of the generated Stokes wave. Finally, a unified realistic model to study the growth of Stokes pulse in short and long pulse regimes through SRS process has been presented in \cite{unified_theory}. Coupled nonlinear equations have been derived describing the evolution of pump and Stokes wave for both the regimes. More recently the role of noise-seeded SRS in degrading the spectral coherence of SC spectrum has been investigated (up to 5 ps)  \cite{Limit_Coherent_SC}.

In this work, we report a statistical analysis describing the evolution of the spectral coherence of noise-seeded Stokes wave in optical fiber for short and long pulse regimes in a unified manner. Our numerical results explore the inquisitive property of Stokes wave in long pulse regime, which reveals a strong synchronization between pump and noise-seeded Stokes pulse inside optical fiber. Phase synchronization is manifested by the gradual stabilization of the spectral coherence of the Stokes wave until it reaches to its maximum value. The paper is organized as follows: In section II and III, we solve coupled nonlinear equation of pump and Stokes wave by split-step Fourier method for short and long pulse regime, respectively. A detailed analysis of the spectral coherence of Stokes wave for the both regions are carried out. Finally conclusion is given in section IV.

\section{Short pulse regime ($<$ 1 ps)}
 The evolution of pump and the Stokes wave in this regime are described by following coupled nonlinear equations \cite{unified_theory}, 
\begin{figure}[t]
\centering
 {\includegraphics[width=8.5 cm,height=9 cm]{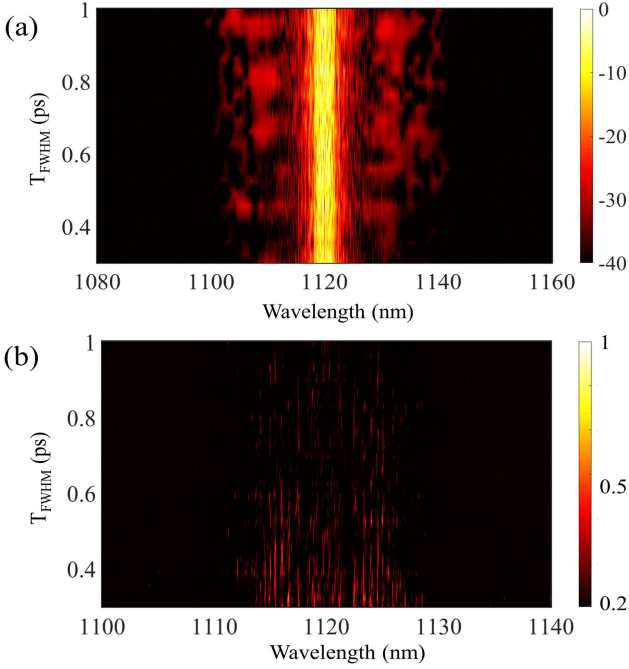}}
\caption{Evolution of (a) spectral intensity (b) spectral coherence of Stokes wave generated by pump power of 1.8 kW after 50 m propagation through fiber as a function of varying pump pulse width.}
\label{fig:pulse_width_var_LPR}
\end{figure}

 \begin{figure}[t]
 \centering
 {\includegraphics[width=8.5 cm,height=9 cm]{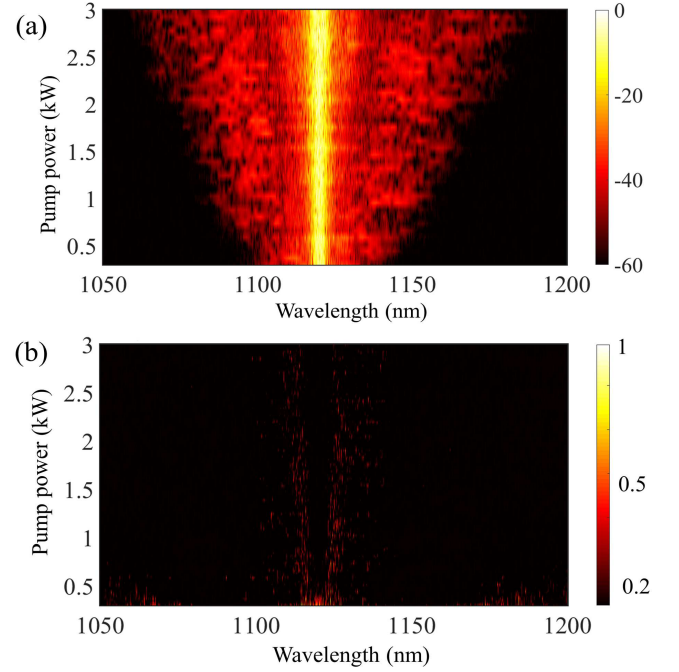}}
 \caption{Evolution of (a) spectral intensity (b) spectral coherence of Stokes wave as a function of input peak pump power for 0.6 ps pulse width at over 50 m of propagation distance.}
 \label{fig:pump_power_var_LPR}
 \end{figure}
 
 \begin{figure}[t]
 \centering
 {\includegraphics[width=8.5 cm,height=9 cm]{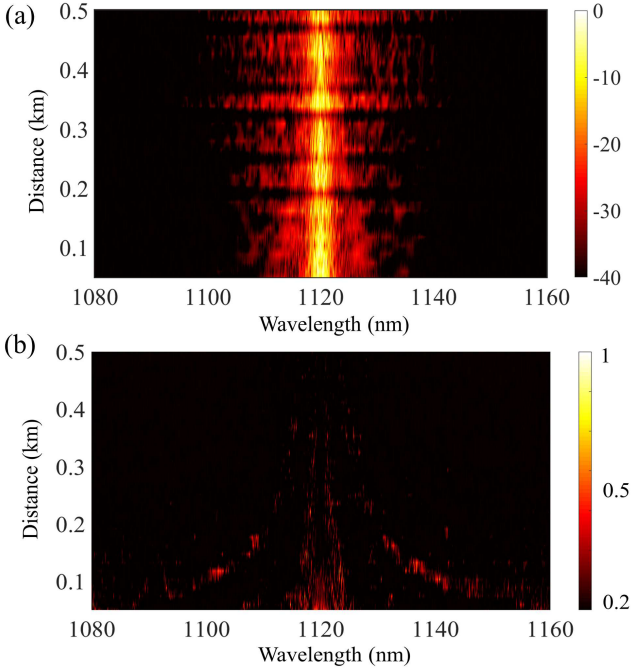}}
 \caption{Evolution of (a) Spectral intensity (b) Spectral coherence of Stokes wave as a function of propagation distance for peak pump power 1.8 kW and pulse width 0.6 ps. }
 \label{fig:length_var_LPR}
 \end{figure} 
 
  \begin{figure*}[t]
   \centering
{\includegraphics[width=18 cm,height=15 cm]{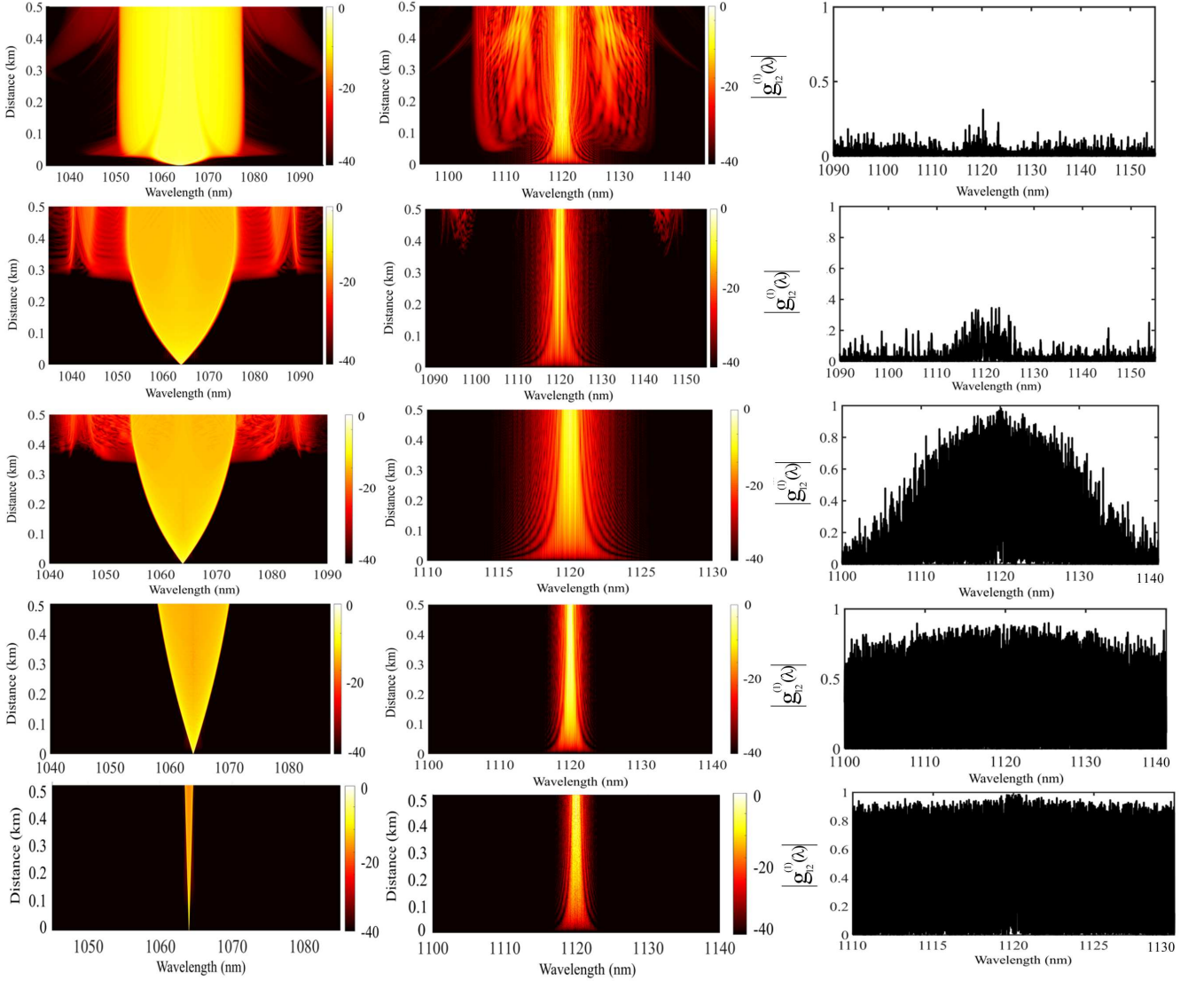}}
   \caption{For pulse width of 5 ps, 50 ps, 100 ps, 250 ps and 500 ps of the pump pulse: (a), (d), (g), (j) and (m), respectively, shows the spectral intensity of pump after .5 km propagation. (b), (e), (h), (k) and (n), respectively, represents the corresponding generated Stokes waves. (c), (f), (i), (l) and (o), respectively, represents the evolution of spectral coherence for varying pulse width.}
   \label{fig:pulse_width_var_HPR}
   \end{figure*}

\begin{equation}
\begin{split}
\MoveEqLeft
\frac{\partial u_{p}}{\partial z}+\frac{1}{v_{gp}}\frac{\partial u_{p}}{\partial t}+i\frac{\beta_{2p}}{2}\frac{\partial^{2}u_{p}}{\partial t^{2}}-\frac{\beta_{3p}}{6}\frac{\partial^{3}u_{p}}{\partial t_{3}}+\frac{\alpha_{p}}{2}u_{p} = \\ & i\gamma_{p}(1-f_{R})u_{p}(|u_{p}|^{2}+2|u_{s}|^{2}) +i\gamma_{p}f_{R}u_{p}\\&\int_{-\infty}^\infty h_{r}(t-t')[|u_{p}(t')|^2+|u_{s}(t')|^2] \,dt'+\\&i\gamma_{p}f_{R}u_{s}\int_{-\infty}^{\infty}h_{r}(t-t')u_{p}(t')u_{s}^{*}(t')exp[i\Omega_{R}(t-t')]\,dt'+\\& iu_{s}\int_{-\infty}^{\infty}h_{p}(t-t')f_{N}(z,t')exp[i\Omega_{R}(t-t')]\,dt'
\label{pump_evol} 
\end{split}
\end{equation}
\begin{equation}
\begin{split}
\MoveEqLeft
\frac{\partial u_{s}}{\partial z}+\frac{1}{v_{gs}}\frac{\partial u_{s}}{\partial t}+i\frac{\beta_{2s}}{2}\frac{\partial^{2}u_{s}}{\partial t^{2}}-\frac{\beta_{3s}}{6}\frac{\partial^{3}u_{s}}{\partial t_{3}}+\frac{\alpha_{s}}{2}u_{s} = \\ & i\gamma_{s}(1-f_{R})u_{s}(|u_{s}|^{2}+2|u_{p}|^{2}) +i\gamma_{p}f_{R}u_{s}\\&\int_{-\infty}^\infty h_{r}(t-t')[|u_{p}(t')|^2+|u_{s}(t')|^2] \,dt'+\\&i\gamma_{s}f_{R}u_{p}\int_{-\infty}^{\infty}h_{r}(t-t')u_{s}(t')u_{p}^{*}(t')exp[i\Omega_{R}(t-t')]\,dt'+\\& iu_{p}\int_{-\infty}^{\infty}h_{s}(t-t')f_{N}(z,t')exp[i\Omega_{R}(t-t')]\,dt'
\label{stoke_evol} 
\end{split}
\end{equation}

In equations (\ref{pump_evol}) and (\ref{stoke_evol}), $\beta_{2j}$ is the group velocity distribution (GVD), $v_{gj}$ is the group velocity, \(u_{j}\) is the field amplitude, which is normalized as,
\begin{equation}
\begin{split}
\MoveEqLeft
u_{j}=\kappa A_{j}[\int_{-\infty}^{\infty}T^{2}(x,y)\,dx\,dy]^{1/2},   (j=p,s)
\label{field_amplitude} 
\end{split}
\end{equation} 
where \textit{T}(x,y) is the electric field distribution perpendicular to the direction of propagation, $\kappa$ is a factor which normalize $u_{j}$ such that $|u_{j}|^2$ represents power and defined as,
\begin{equation}
\kappa=\frac{1}{2}n_{j}(\epsilon_{0}/\mu_{0})
\end{equation}
where $n_{j}$ is the linear index of the fiber. $\gamma_{j}$ is the nonlinear coefficient and defined as,
\begin{equation}
\begin{split}
\MoveEqLeft
\gamma_{j}=\frac{\omega_{j}n_{2}}{c A_{eff} \kappa^{2}},  n_2=\dfrac{3}{8n_{j}}\chi_{K}(1+\frac{2\chi_{0}}{3\chi_{K}}), \\&
A_{eff}=\dfrac{(\int_{-\infty}^{\infty}\int_{-\infty}^{\infty} T^{2}\,dx\,dy)^{2}}{\int_{-\infty}^{\infty}\int_{-\infty}^{\infty} T^{4}\,dx\,dy}, \chi_{R}(t)=\chi_{0}h_{r}(t) 
\end{split}
\end{equation} 
where $n_{2}$ is the nonlinear refractive index, c is the velocity of light in vacuum, $A_{eff}$ is the effective mode area, $\chi_{K}$ is the Kerr susceptibility, $\chi_{R}$ is the third order time dependent nonlinear susceptibility leading to Raman scattering process and $\chi_{0}$ is the peak value of $\chi_{R}(t)$, $f_{R}$ is the fractional contribution of nuclei to the total nonlinear polarization, $h_{j}$ is the response function for noise term, defined as,
\begin{equation}
h_{j}(t)=\frac{\omega_{j} R_{N}(t)}{4c n_{j}}
\end{equation}
where $R_{N}(t)$ is the response function that accounts spontaneous scattering during propagation and convert a Langevin noise source function $F_{N}$ which incorporates the random vibration of the silica molecule due to temperature \cite{Raman_generation2}, into susceptibility $\chi_{N}$,
 \begin{equation}
 \begin{split}
\chi_{N}=\int_{-\infty}^{\infty}R_{N}(t-t')F_{N}(t')\,dt'
 \end{split}
 \end{equation}
The noise force $F_{N}$ can be expressed as,
 \begin{equation}
 \begin{split}
F_{N}(z,t)=\frac{1}{2}\widehat{x}[f_{N}(z,t)exp(-i\Omega_{R}t)+c.c]
 \end{split}
 \end{equation}
where $f_{N}$ is the slowly varying function of random force and $\Omega_{R}=\omega_{p}-\omega_{s}$ is the difference between carrier frequency of the pump and stokes wave, respectively.

To interpret equations (\ref{pump_evol}) and (\ref{stoke_evol}), the first term on the left-hand side of the both equation represents for the pulse envelope change with distance whereas the next three terms accounts the effect of group velocity, GVD and the third-order dispersion, respectively. The last terms of the left-hand side represents the fiber loss for both pulses. The first two terms on the right-hand side account the effect of SPM and XPM. The next two terms represent the molecular contribution of SPM and XPM, respectively and it also accounts the effect of interpulse SFS and intrapulse CFS.
The last term corresponds for the spontaneous Raman scattering process.
To investigate the evolution of pump and the Stokes wave, we solve eq. (\ref{pump_evol}) and eq. (\ref{stoke_evol})  using split-step Fourier method. The noise, which occurs due to thermal and quantum fluctuations of vibrational mode through spontaneous Raman scattering process, is modeled as a Markovian stochastic process with Gaussian statistics of zero mean and unit variance i.e. $<f_{N}(z,t)>=0$ \cite{ASE_noise}. Due to the shot noise, the input pulse experience fluctuations in power density and random phase of photon ranging from 0 to 2$\pi$. This has been included in simulation. The input pump pulse has the form of, \(u_{p}(0,t)=\sqrt{P_{0p}} sech(t/t_{p})\), where $P_{0p}$ is the pump power and $t_{p}$ is the pump pulse width. Initially, Stokes pulse is taken as white Gaussian noise distribution. 
To investigate coherence property of the generated Stokes wave, an ensemble of generated spectrum is simulated with varying input conditions. The input pump pulse is added with quantum shot noise where one photon per mode with random phase on each spectral discretization bin. An ensemble of 40 simulations  are performed considering same identical input condition with different stochastic noise on the input laser pump pulse. The shot-to-shot fluctuation describing modulus of first-order coherence at zero path difference is given by \cite{dudley_supercontinuum},
 \begin{equation}
 \begin{split}
|g_{12}^{(1)}(\lambda)|=\mid \frac{<E_{1}^{*}(\lambda)E_{2}(\lambda)>}{\sqrt{<|E_{1}(\lambda)|^{2}><|E_{2}(\lambda)|^{2}>}} \mid
 \end{split}
 \label{coherence}
 \end{equation}

The angular bracket represents the ensemble average of the adjacent pair $[E_{1}(\lambda),E_{2}(\lambda)]$ of generated spectrum from independent simulations. The degree of spectral coherence varies between 0$\leq$ $|g_{12}^{(1)}|$ $\leq$1 where 0 corresponds to incoherent spectrum representing high shot-to-shot fluctuation, whereas 1 corresponds to coherent spectrum representing perfect stability in phase and amplitude.
We solve equation (\ref{coherence}) numerically several times to obtain the statistical evolution of coherence of generated Stokes wave. Parameters are taken from ref. \cite{Raman_generation2}. The pump is Gaussian shaped pulse at 1064 nm. Both the Stokes pulse at 1120 nm and the pump fall in the normal dispersion region of the fiber. Fiber parameters for the pump wavelength are $\beta_{2p}$=25 $ps^{2}$/km, $\gamma_{p}$=5.06 $w^{-1}km^{-1}$ and $g_{p}$=2.34 $w^{-1} km^{-1}$. The value of these parameters for the Stokes pulse are scaled by the factor $\lambda_{p}/\lambda_{s}$=0.95. The walk-off length \textit{d} is 2.2 ps/m, where \textit{d} is given by, 
\begin{equation}
 \begin{split}
d=\dfrac{1}{v_{gp}}-\dfrac{1}{v_{gs}}
 \end{split}
 \end{equation}
 
The numerical code solving equation (\ref{pump_evol}) and (\ref{stoke_evol}) is repeated by 40 times with the above parameters. In every simulation, fluctuating noise in the input leads to significant difference in Stokes properties. The statistical analysis of the evolution of coherence of Stokes with pulse width variation, input pump power and fiber length is performed from the simulated ensemble.
Fig. \ref{fig:pulse_width_var_LPR}(a) and Fig. \ref{fig:pulse_width_var_LPR}(b) exhibit the spectral evolution and the corresponding spectral coherence, respectively, for the fiber length of 50 m. Very low spectral coherence is observed in this region which supports the results as discussed in \cite{Limit_Coherent_SC}. The amplification of noise causes high pulse-to-pulse fluctuation, resulting into poor coherence. As the pump pulse duration increases (fs to ps), SRS becomes more dominant and exhibits large fluctuation in phase and amplitude, leading to degradation of coherence as shown in Fig. \ref{fig:pulse_width_var_LPR}(b).

To investigate the effect of input pump power on the coherence of the Stokes wave, the pump energy is varied. Fig. \ref{fig:pump_power_var_LPR}(a) and Fig. \ref{fig:pump_power_var_LPR}(b) depict the evolution of spectral intensity and spectral coherence of Stokes wave, respectively, over the propagation distance of 50 m long the fiber. It is observed that with the increase in power, Stokes spectrum also broadens but there is progressively degradation of spectral coherence.
 To carry out the effect of fiber length on the coherence of Stokes wave, code runs for fixed pump power at 1.8 kW and pulse width at 0.6 ps. Fig. \ref{fig:length_var_LPR}(a) and Fig. \ref{fig:length_var_LPR}(b) show the spectral evolution and corresponding coherence of Stokes wave, respectively. It is observed that Stokes spectrum gets compressed as it propagates through fiber. This is due to the narrowing of Raman gain with propagation. The central frequency of the Stokes wave are amplified over wings, which causes narrowing of Stokes spectrum \cite{Raman_generation2}. The coherence also gets gradually degrades over distance. 
  The incoherent spectral evolution of noised-seeded Stokes wave are discussed in this section. In high pulse region the scenario is totally different which we discuss in the next section. 
  
 \section{Long pulse regime ($>$ 1 ps)}
 The evolution of pump and Stokes waves governed by equations (\ref{pump_evol}) and (\ref{stoke_evol}), relies on slowly varying envelope approximation. These equations are no longer valid for pulse width greater than 1 ps. For broad pulse width, the pulse envelope can be treated as constant in comparison with the time scale of Raman response function $h(r)$. The simplified form of coupled nonlinear equation governing pump and Stokes evolution in long pulse regime can be given by \cite{unified_theory},

 \begin{equation}
 \begin{split}
 \MoveEqLeft
 \frac{\partial u_{p}}{\partial z}+\frac{1}{v_{gp}}\frac{\partial u_{p}}{\partial t}+i\frac{\beta_{2p}}{2}\frac{\partial^{2}u_{p}}{\partial t^{2}}-\frac{\beta_{3p}}{6}\frac{\partial^{3}u_{p}}{\partial t_{3}}+\frac{\alpha_{p}}{2}u_{p} = \\ & i\gamma_{p}u_{p}[|u_{p}|^{2}+(2-f_{R})|u_{s}|^{2}] +i/2u_{p}u_{s}\\&\int_{-\infty}^\infty g_{p}(t-t')u_{s}^{*}(t') exp[i\Omega_{R}(t-t')] \,dt'+\\& iu_{s}\int_{-\infty}^{\infty}h_{p}(t-t')f_{N}(z,t')exp[i\Omega_{R}(t-t')]\,dt'
 \label{pump_evol_LPR} 
 \end{split}
 \end{equation}
  \begin{equation}
 \begin{split}
 \MoveEqLeft
 \frac{\partial u_{s}}{\partial z}+\frac{1}{v_{gs}}\frac{\partial u_{s}}{\partial t}+i\frac{\beta_{2s}}{2}\frac{\partial^{2}u_{s}}{\partial t^{2}}-\frac{\beta_{3s}}{6}\frac{\partial^{3}u_{s}}{\partial t_{3}}+\frac{\alpha_{s}}{2}u_{s} = \\ & i\gamma_{s}u_{s}[|u_{s}|^{2}+(2-f_{R})|u_{p}|^{2}] +i/2|u_{p}|^{2}\\&\int_{-\infty}^\infty g_{s}(t-t')u_{s}(t') exp[-i\Omega_{R}(t-t')] \,dt'+\\& iu_{p}\int_{-\infty}^{\infty}h_{s}(t-t')f_{N}^{*}(z,t')exp[-i\Omega_{R}(t-t')]\,dt'
 \label{stoke_evol_LPR} 
 \end{split}
 \end{equation}
 where Raman gain coefficient $g_{j}$ can be defined as, 
\begin{equation}
 \begin{split}
g_{j}=2f_{R}\gamma_{j}|\widetilde{h_{r}^{''}}(\Omega_{R})|   (j=p,s)
 \end{split}
 \end{equation}
 $h_{r}(\Omega_{R})$ is the Raman response function at frequency $\omega$=$\Omega_{R}$. The real [$\widetilde{h_{r}^{'}}(\omega_{R})$]  and imaginary parts $[\widetilde{h_{r}^{''}}(\omega_{R}))]$ of $h_{r}(\omega_{R})$ stand for Raman induced index change and Raman gain, respectively.
 
In order to investigate the evolution of spectral coherence at long-pulse regime, equations (\ref{pump_evol_LPR}) and (\ref{stoke_evol_LPR}) are solved repeatedly. The first order mutual coherence has been calculated using equation (\ref{coherence}). The fiber parameters are similar to the case for short-pulse regime. Fig. \ref{fig:pulse_width_var_HPR} represents the simulated results for the evolution of pump pulse, Stokes pulse and the spectral coherence of the Stokes pulse for gradually varying pump pulse width (5 ps, 50 ps, 100 ps, 250 ps and 500 ps). It is observed that with the increase in pulse width, the Stokes wave which builds from the random noise, gradually develops towards high spectral coherence and reaches to its maximum value. In order to quantify the overall spectral coherence across the spectrum, we have calculated the spectrally averaged coherence (SAC) which is given by \cite{dudley_coherence},

\begin{equation}
 \begin{split}
<|g_{12}^{(1)}(\lambda)|>=\mid \frac{\int |g_{12}^{(1)}(\lambda, 0)||E(\lambda)|^2 d\lambda }{\int |E(\lambda)|^2 d\lambda } \mid
 \end{split}
 \label{overall_coherence}
 \end{equation}
 
 \begin{figure}[t]
 \centering
 {\includegraphics[width=8.5 cm,height=4.5 cm]{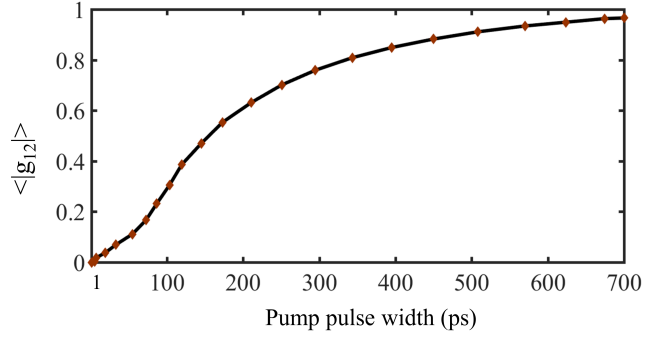}}
 \caption{Evolution of overall spectral coherence with pump pulse width.}
 \label{overall spectral coherence}
 \end{figure} 
The value of SAC relies between 0 and 1. Fig. \ref{overall spectral coherence} shows the variation of SAC of the Stokes spectrum with pump pulse width in the long pulse regime which is calculated over the bandwidth (1110-1140 nm). The simulated results reveal that the phase synchronization of the Stokes wave where the degree of spectral coherence of the Stokes wave gradually stabilize with the increase in pulse width and beyond 500 ps pulse width, very high spectral coherence is obtained. The origin of the phase synchronization can be understood in terms of walk-off length. With the increase in pump pulse width, the corresponding walk-off length also increases as the walk-off length is proportional to the pulse width. The enhanced walk-off length allows the pump and the noise-seeded Stokes pulse to propagate together and hence, pump pulse can trigger the Stokes pulse for long path. As a result, the initial random phase of the Stokes wave tends to stabilize gradually to its maximum value with the increase in pulse width and consequently high spectral coherence is achieved. 

\section{Conclusion}
We have investigated the detailed statistical evolution of spectral coherence of the noise-seeded Stokes wave in short and long pulse regimes in a unified manner. Simulation results largely reveal the undisclosed coherence properties of the Stokes wave in long pulse regime. Noise-seeded Stokes pulse which offers poor
spectral coherence in short pulse regime, exhibits gradual increment of spectral coherence to its maximum value with the increase in pulse width in long pulse regime. The findings would enhance the prolonged ideas and will provide new window for the realization of cascaded Raman based coherent broadband spectrum, coherent Raman laser and so on.


\end{document}